\documentstyle[preprint,aps,epsf,floats]{revtex}
\begin{document}
\tighten

\def\bfl{{\bbox \ell}}
\def\bull{\vrule height .9ex width .8ex depth -.1ex}
\def\MeV{{\rm MeV}}
\def\GeV{{\rm GeV}}
\def\Tr{{\rm Tr\,}}
\def\nrcpt{NR\raise.4ex\hbox{$\chi$}PT\ }
\def\ket#1{\vert#1\rangle}
\def\bra#1{\langle#1\vert}
\def\ltap{\ \raise.3ex\hbox{$<$\kern-.75em\lower1ex\hbox{$\sim$}}\ }
\def\gtap{\ \raise.3ex\hbox{$>$\kern-.75em\lower1ex\hbox{$\sim$}}\ }
\newcommand{\gsim}{\raisebox{-0.7ex}{$\stackrel{\textstyle >}{\sim}$ }}
\newcommand{\lsim}{\raisebox{-0.7ex}{$\stackrel{\textstyle <}{\sim}$ }}

\def\Journal#1#2#3#4{{#1} {\bf #2}, #3 (#4)}

\def\NCA{\em Nuovo Cimento}
\def\NIM{\em Nucl. Instrum. Methods}
\def\NIMA{{\em Nucl. Instrum. Methods} A}
\def\NPB{{\em Nucl. Phys.} B}
\def\NPA{{\em Nucl. Phys.} A}
\def\PLB{{\em Phys. Lett.}  B}
\def\PRL{\em Phys. Rev. Lett.}
\def\PRD{{\em Phys. Rev.} D}
\def\PRC{{\em Phys. Rev.} C}
\def\PRA{{\em Phys. Rev.} A}
\def\PR{{\em Phys. Rev.} }
\def\ZPC{{\em Z. Phys.} C}
\def\PREP{{\em Phys. Rep.}  }
\def\ANN{{\em Ann. Phys.} }
\def\SCI{{\em Science} }
\def\CJP{{\em Can. J. Phys.}}

\def\a{{\alpha} }
\def\b{{\beta} }
\def\c{{\gamma} }

\def\e{{\epsilon} }

\def\s2{{\sigma_2} }

\def\D{{\Delta} }
\def\dfourp{{\int \frac{d^4p}{(2 \pi)^4}} }
\def\dthreep{{\int \frac{d^3p}{(2 \pi)^3}} } 
\preprint{\vbox{
\hbox{40561-75-INT99}
}}
\bigskip
\bigskip

\title{Color Superconductivity in Asymmetric Matter
}
\author{Paulo F. Bedaque}
\address{Institute for Nuclear Theory, University of Washington,
Seattle, WA 98195-1550
\\ {\tt bedaque@phys.washington.edu}}

\maketitle

\begin{abstract}
The influence of different chemical potential for different flavors
on color superconductivity is analyzed.
 It is found that there is a first order transition
as the asymmetry grows.
This transition proceeds through the formation of bubbles  of low density, 
flavor asymmetric normal phase inside a high density, superconducting phase 
with a gap {\it larger} than the one found in the symmetric case. 
For small fixed asymmetries the system is normal at low densities and superconducting
only above some critical  density. For larger asymmetries the two massless quarks 
system stays in the
mixed state for arbitrarily high densities.
\end{abstract}

\vfill\eject

A lot of progress has been made recently in mapping the phase diagram of QCD
in the low temperature, high baryon density region.
The main idea advanced by this research is the almost unavoidable 
formation of quark-quark condensates (pairing)  
and the spontaneous breaking (or restoration) of color-flavor 
symmetries of QCD \cite{Barrois,BailinLove}.
These ideas may have an impact on  the phenomenology of neutron stars and of heavy 
ion collisions at SPS energies where compressed strong interacting matter is believed
to exist.

Since lattice calculations at finite density are unfeasible at this date two 
alternative approaches
are used to study this problem. One, valid at asymptotically high densities, 
is the use of perturbation theory. The gaps due to perturbative QCD were believed to 
be small\cite{BailinLove}, exponentially suppressed as in weak coupling BCS theory. 
It was 
recently realized\cite{PisarskiRischke,Son,Hong,SW3,PR,rockefeller,Hsu2}, 
however, that the main pairing interaction in this regime is 
due to the 
exchange of magnetic gluons. These interactions make the normal phase even more unstable
than short range forces do and result on a parametrically much larger gap. Extrapolating
this down to more accessible densities one could expect gaps  up to $100$ MeV at
chemical potentials  $\mu\sim 1000$ MeV. It remains to be seen 
at which density the sub-leading correction (in powers of the coupling) become
large and invalidate the perturbative expansion.
Another approach to the problem is the use of phenomenological models such as
NJL-like models and the instanton liquid model
\cite{ARW1,NJL,steph,RappETC,RappETC2,bergesraj,stephetal,CarterDiakonov}. 
In particular, a large class of four-fermion interaction models   
was  considered. Fortunately, for semi-quantitative purposes, they tend to give
very similar results. This is due to the fact that the formation of the condensate
depends only on the most attractive  channel between quarks at the Fermi surface.
Different four-fermion interactions that are both attractive in the scalar-isoscalar
color ${\bar 3}$
channel and of roughly the same strength produce similar BCS instabilities. 
Renormalization group analysis support this conclusion and help in classifying
all four-fermion operators according to their importance to the formation of the gap
\cite{Hsu1,SW0}.

With the help of these phenomenological models the most likely scenarios of 
symmetry breaking were identified\cite{Barrois,BailinLove,ARW1,RappETC}.
In the case of $2$ massless quarks chiral symmetry is likely restored at some
critical density and color 
$SU_c(3)$ is partially broken down to $SU_c(2)$ by the formation of the condensate (2SC)
\begin{equation}
\langle q^{\a a}_{L i}  q^{\b b}_{L j} \rangle =
\langle q^{\a a}_{R i}  q^{\b b}_{R j} \rangle
\sim \e^{\a\b 3} \e^{a b}\e_{i j} 
\label{eq:2SCcondensate}
\end{equation}
\noindent
where the  $\a,\b,...$ are color indices, $a,b,...$ are flavor indices,
$i,j,...$ are spinor indices and  $q_{L,R}$
are left (right) handed spinors.
The condensate in eq.~(\ref{eq:2SCcondensate}) singles out the third color 
(green from now on), only blue and red quarks pair up.

In the three massless quarks case the symmetry breaking pattern is more subtle
\cite{CFL,SW1}.
The preferred condensate is the so-caled color-flavor locked (CFL) state
\begin{equation}
\langle q^{\a a}_{L i} \sigma_{2} q^{\b b}_{L j} \rangle =
-\langle q^{\a a}_{R i} \sigma_{2} q^{\b b}_{R j} \rangle
\sim \e_{i j} (\kappa_1 \delta_{\a a} \delta_{\b b} 
             +  \kappa_2 \delta_{\a b} \delta_{\b a}) 
\label{eq:CFLcondensate}
\end{equation}
\noindent
The condensate eq.~(\ref{eq:CFLcondensate})
breaks $SU_c(3)\times SU_L(3)\times SU_R(3)$ down to 
$SU_{c+L+R}(3)$, leaving  
the ground state invariant under a combined
color-flavor rotation. The remaining symmetry $SU_{c+L+R}(3)$ is the same as
the one of the vacuum after spontaneous breaking of 
chiral symmetry ($SU_V(3)$). This suggests that there 
is smooth connection between the hadronic phase to the CFL phase.
The spectrum of excitations is identical in both cases: the gluon octet
is mapped onto the vector meson octet, the quarks onto the baryon octet and 
the $8$ Goldstone bosons onto the pseudo-scalar mesons.

The real world lies between the $2$ and $3$  massless quark case since
the mass $m_s$ of the strange quark is neither much larger nor much smaller
than the QCD scale. At arbitrarily high densities the presence of the strange 
quark mass is irrelevant but not at lower densities. It is difficult then to 
determine 
whether the CFL phase extends all the way to lower densities and connects to
the hadronic phase or whether there is a window in density with the 2SC phase.
Another difference between the discussion sketched above and the potential 
real world application
is the difference in densities among the different quark flavors. 
In heavy ion collisions
there is an excess of  $\approx 15\%$ in the number of down quarks 
(minus down anti-quarks) relative to the 
up quarks, and zero net strangeness at all. There are also significant asymmetries
in neutron star cores.

Here we will analyze the two massless quarks phase
in a model that conserves quark flavors
 
We choose our 
interaction to be
a contact four fermion coupling with spin, flavor and color structure abstracted from
the one gluons exchange projected on the color ${\bf \bar 3}$ channel (the ${\bf 6}$
is repulsive). 
Our starting point is then the Lagrangian
\begin{equation}
{\cal L-\mu}_a {\cal N}_a=\psi^\dagger_{\a a} 
((i\sigma^\mu \partial_\mu + \mu)\delta_{a b}+ \delta \tau^3_{a b})
\delta_{\a \b}\psi_{\b b}
-g^2 \psi^T_{\a a} \sigma_2 \psi^\dagger_{\b b} 
\psi^\dagger_{\c b} \sigma_2 \psi^*_{\delta a}
(\delta_{\a\c}\delta_{\b\delta}- \delta_{\a\delta}\delta_{\b\c}),
\label{eq:L}
\end{equation}
where $\mu=(\mu_u+\mu_d)/2$ is the chemical potential averaged over flavors
(equal to one third of the baryon chemical potential), $\delta=(\mu_u-\mu_d)/2$
is their difference, ${\cal N}_a$ is the operator that counts the total
amount of flavor $a$ (quarks minus antiquarks) and the spinor indices were omitted.
The interaction in eq.~(\ref{eq:L}) does not mix left and right handed quarks
and a sum over handedness is implicit throughout the paper. The matrices $\sigma^\mu$
are defined as $\sigma^\mu=(1,-\vec{\sigma})$ for the left handed quarks and
$\sigma^\mu=(1,\vec{\sigma})$ for the right handed ones.
The mean field approximation, although not arising as a  systematic expansion
in a small parameter is usually reliable to determine the phase structure of the theory.
It is exact in the limit of large number of flavors (keeping $g^2 N_F$ constant)
and on the weak coupling limit.
To perform the mean field approximation we rewrite the model defined by 
eq.~(\ref{eq:L}) by
introducing a dummy field $\D_{\a a \b b}$
\begin{eqnarray}
{\cal L}+\mu_a{\cal N}_a&=&\psi^\dagger_{\a a} 
((i\sigma^\mu \partial_\mu + \mu)\delta_{a b}
+ \delta \tau^3_{a b})\delta_{\a \b}\psi_{\b b}
+\frac{1}{g^2}\D_{\a a \b b}\D^*_{\c b \delta a}\e_{\a\b\c\delta}\nonumber\\ 
&+&\psi^T_{\a a} \sigma_2 \psi^\dagger_{\b b}\D^*_{\c b \delta a}\e_{\a\b\c\delta}
+\psi^\dagger_{\c b} \sigma_2 \psi^*_{\delta a}\D_{\a a \b b}\e_{\a\b\c\delta},
\label{eq:L2}
\end{eqnarray}
where $\e_{\a\b\c\delta}=\delta_{\a\c}\delta_{\b\delta}- 
\delta_{\a\delta}\delta_{\b\c}$. 
It is convenient to use the Nambu-Gorkov formalism and write eq.~(\ref{eq:L2}) as
\begin{equation}
{\cal L+\mu}_a {\cal N}_a=
 \left(\begin{array}{c c} 
         \psi^\dagger    &    \psi^T \sigma_2\end{array} \right)
{\cal M}
 \left(\begin{array}{c} \psi\\
                      \sigma_2\psi^*\end{array}\right)
+\frac{1}{g^2}\D_{\a a \b b}\D^*_{\c b \delta a}\e_{\a\b\c\delta},
\label{eq:L3}
\end{equation} 
with
\begin{equation}
{\cal M}= \left(\begin{array}{c c} 
                p_\mu\sigma^\mu+\mu + \delta \tau^3 & \Phi\\ 
                \Phi^*                &  p_\mu\bar\sigma^\mu-\mu + \delta \tau^3
               \end{array}\right),
\label{eq:M}
\end{equation} 
where $\bar\sigma^\mu=(1,\vec{\sigma})$ 
for the left handed quarks
and $\bar\sigma^\mu=(1,-\vec{\sigma})$ 
for the right handed ones. The matrix
$\tau^3$ acts on flavor space (and is diagonal in color and spin)
and $\Phi$ is the operator 
\begin{equation}
\Phi_{\a a \b b}=\D_{\c a \delta b}\ \e_{\c\delta\a\b}, 
\end{equation} 
\noindent
acting on color-flavor space.
 
The mean field approximation 
amounts to estimating the path integral by its saddle value point. 
Ignoring then the
fluctuations on the $\D_{\a b \b b}$ field we are left with a Gaussian 
integral over the fermion fields that produces the familiar 
${\rm det}({\cal M})={\rm Tr}\ {\rm log}({\cal M})$ term. 
The potential $\Omega(\D,\mu,\delta)$
defined by
\begin{equation}
e^{-i\Omega(\D,\mu,\delta)}=\int {\cal D}\psi{\cal D}\psi^\dagger 
e^{i\int {\cal L}+\mu_a {\cal N}_a} 
\end{equation}
\noindent 
is the expectation value of the Hamiltonian ${\cal H}-\mu_a{\cal N}_a$
in the state that minimizes ${\cal H}-\mu_a {\cal N}_a$ in the subspace of states with
a given value of  $\D$.
In our case 
\begin{equation}
\Omega(\D,\mu,\delta)=\frac{i}{2}\dfourp {\rm tr\ log}{\cal M}
             +\frac{2}{g^2}\D_{\a a \b b}\D^*_{\c b \delta a}\e_{\a\b\c\delta},  
\end{equation}
where the trace is over spin, color and flavor spaces. 
The value of $\D_{\a a \b b}$ in the ground is obtained by minimizing
$\Omega(\D,\mu,\delta)$ at fixed $\mu$. We make an ansatz, 
corresponding to the 2SC state,
for the form of the order parameter
\begin{equation}
\D_{\a a \b b}=\frac{\D}{2}  i\e_{ ab} \e_{\a \b 3}.
\end{equation}
\noindent
As mentioned before, this choice breaks color $S_c(3)$ down to $S_c(2)$.
By using a basis in
spin space that diagonalizes $\vec{p}.\vec{\sigma}$ and by doing the 
similarity transformation
\begin{equation}
{\cal M}\rightarrow U {\cal M} U^{-1},\ \ \  U= \left(\begin{array}{c c} 
                                                   1 & 0\\ 
                                                   0 & \tau_2
                                                \end{array}\right)
\end{equation}
\noindent
we are left with the task of computing the determinant of
\begin{equation}
{\cal M}=\left(\begin{array}{c c} 
                p_0-\e^{\pm}(p) + \delta \tau^3 & \bar\Phi\\ 
                    \bar\Phi^*          & p_0+\e^{\pm}(p) - \delta \tau^3
               \end{array}\right),
\label{eq:M2}
\end{equation}
\noindent
where
\begin{equation}
\bar\Phi_{\a\b}=-i \D \e_{\a\b 3}
\label{eq:phibar}
\end{equation}
\noindent
acts only in color space and $\e^\pm(p)=\pm p-\mu$. Using the relations
\begin{equation}
{\rm tr\  log} \left(\begin{array}{c c}A & B\\
                                   C & D
                 \end{array}\right)  =
{\rm Tr\ log}(-BC+BDB^{-1}A),
\label{eq:tommyboy}
\end{equation}
\noindent

\begin{equation}
\bar\Phi^2=-P,
\end{equation}
\noindent
where $P$ is the projector on the first two colors subspace, and
\begin{eqnarray}
{\rm Tr}_{\rm flavor} \log(a+b \tau^3)&=&\log(a+b)+\log(a-b)\\
{\rm Tr}_{\rm color} \log(a+b  P)&=&(N_c-1)\log(a+b)+\log(a),
\end{eqnarray}
we arrive at
\begin{eqnarray}
\Omega(\D,\mu,\delta)&=\frac{\D^2}{g^2}+\sum_{+,-}
\frac{1}{2}\dfourp & [  2 \log((p_0+\delta)^2-E^{\pm 2}(p))
                                  +  \log((p_0+\delta)^2-\e^{\pm 2}(p))
                                       \nonumber\\
                  & &+ 2 \log((p_0-\delta)^2-E^{\pm 2}(p))
                                  +\log((p_0-\delta)^2-\e^{\pm 2}(p))]\nonumber\\
                  & &+ C,
\end{eqnarray}
\noindent
where $E^{\pm}(p)=\sqrt{\D^2+\e^{\pm 2}(p)}$ is the energy of the quasi-particle
excitations.
Performing the $p_0$ integration and fixing the constant $C$ by demanding
$\Omega(\D,\mu,\delta)$ has the correct (free theory) value at $\D=0$ we arrive at
\begin{eqnarray}
\Omega(\D,\mu,\delta)&=\frac{\D^2}{g^2}-\frac{1}{2}\dthreep \sum_{+,-}[
      & 2(E^{\pm }(p)-p-\theta(\delta-E^{\pm }(p))(E^{\pm }(p)-\delta))\nonumber\\
      && + |\e^{\pm }(p)|-p-\theta(\delta-|\e^{\pm }(p)|)
                 (|\e^{\pm }(p)|-\delta)]\nonumber\\ 
      && +3 p +\theta(\delta+\e^-(p))(\delta+\e^-(p)).
\label{eq:Omega}      
\end{eqnarray}
\noindent
The first line corresponds to the contribution of the red and blue quarks, the second
to the unpaired green quarks and the last one to anti-particles that can be created when
$\delta<\mu$. The mean particle number density of each flavor is obtained by
\begin{eqnarray}
n_u+n_d&=&-\frac{\partial}{\partial\mu}\Omega(\D,\mu,\delta)\nonumber\\
       &= & \dthreep \sum_{+,-}[
                       -2\frac{\e^\pm(p)}{E^\pm(p)}
                       +2 \theta(\delta-E^\pm(p))\frac{\e^\pm(p)}{E^\pm(p)}\nonumber\\
                      & & -{\rm sgn}(\e^\pm(p))
                       + \theta(\delta-|\e^\pm(p)|) {\rm sgn}(\e^\pm(p))]
                       + \theta(\delta+\e^-(p))\nonumber\\
n_u-n_d&=&-\frac{\partial}{\partial\delta}\Omega(\D,\mu,\delta)\nonumber\\
       &= & \dthreep \sum_{+,-}\left[2\theta(\delta-E^\pm(p))
                               +\theta(\delta-|\e^\pm(p)|)
                               - \theta(\delta+\e^-(p))   \right].
\label{eq:N}      
\end{eqnarray}

\begin{figure}[t]
\centerline{\epsfxsize=4.0in \epsfbox{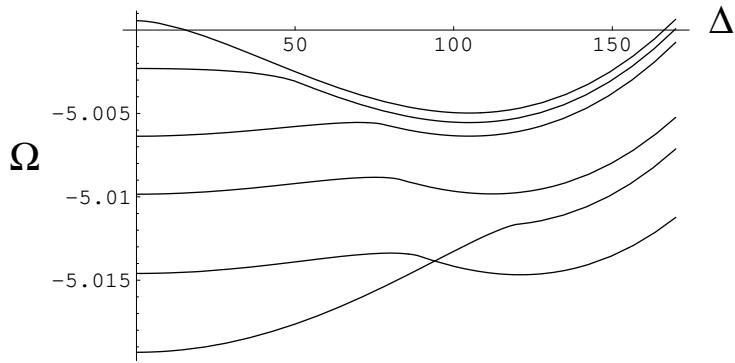}}
\noindent
\caption{\it 
  Potential $\Omega$ in units of $10^{10}$ MeV$^4$ as a function
  of the gap $\D$ in MeV.
  By the order the curves meet the vertical axis, from top to bottom:
i)   $\mu=300$ MeV, $\delta=0$, 
ii)  $\mu=300$ MeV, $\delta=40$ MeV, 
iii) $\mu=300$ MeV, $\delta=77.5$ MeV,
     $(n_u+n_d)_{normal}=3.1\ 10^6 $ MeV$^3,(n_u+n_d)_{super}= 3.91\ 10^6$ MeV$^3$, 
     $(n_u-n_d)_{normal}=2.34\ 10^6$ MeV$^3,(n_u-n_d)_{super}= 0.90\ 10^6$ MeV$^3$, 
iv)  $\mu=310$ MeV, $\delta=83$ MeV,
     $(n_u+n_d)_{normal}=4.31\ 10^6$ MeV$^3,(n_u+n_d)_{super}= 3.45\ 10^6$ MeV$^3$, 
     $(n_u-n_d)_{normal}=1.04\ 10^6$ MeV$^3,(n_u-n_d)_{super}= 2.69\ 10^6$ MeV$^3$, 
v)   $\mu=322.5$ MeV, $\delta=89$ MeV,
     $(n_u+n_d)_{normal}=4.92\ 10^6$ MeV$^3,(n_u+n_d)_{super}= 3.91\ 10^6$ MeV$^3$, 
     $(n_u-n_d)_{normal}=1.22\ 10^6$ MeV$^3,(n_u-n_d)_{super}= 3.14\ 10^6$ MeV$^3$, 
vi)  $\mu=310$ MeV, $\delta=120$ MeV. }  
\label{fig:fig1}
\vskip .2in
\end{figure}
All three dimensional  integrals are cutoff at momenta $\Lambda$.
For illustration, we choose the values $\Lambda=900$ MeV, $g^2/\Lambda^2=4.4$.
They produce a gap of around $\D\sim100$ MeV at $\mu=300$ MeV and $\delta=0$.
In fig. (1) we have $\Omega(\D,\mu,\delta)$ as a function of $\D$ for 
different values of
$\mu$ and $\delta$.  They all correspond to states with the same value of $n_u+n_d$, 
but increasing values of the asymmetry $n_u-n_d$. The minimum of the first
curve, top to bottom,   corresponds to the flavor symmetric, superconducting state.
As mentioned above, the superconducting state consists of paired blue and red 
quarks of both flavors, 
with green quarks left unpaired. 

There are two ways of looking at flavor asymmetric matter: 
one may fix different densities or different chemical potentials. 
Let us first consider the first possibility and imagine increasing the
 asymmetry $n_u-n_d$ 
from its the initial zero value while keeping the total density $n_u+n_d$ fixed.
States with small $n_u-n_d$ values can be created  by 
adding to this symmetric state up-quasi-particles and  down-quasi-holes. 
In the grand canonical formalism used here this is accomplished by 
 increasing the value of $\delta$ from zero to some small value. Since there is
a gap in the spectrum of blue and red quasi-particle, no quasi-particle with these
 colors are created for $\delta<\D$. The flavor asymmetry is made up only with green,
up-quasi-particles and green, down quasi-holes. If all colors were paired we would see
no change in $\Omega$ at all in the superconducting phase until the split between 
chemical potentials
 $\delta$ equaled the gap $\D$. The potential can be more easily lowered in the 
unstable state with $\D=0$ since there quarks of all colors can be created, 
 even with small $\delta$. As a result $\Omega(\D=0,\mu,\delta)$ decreases, 
with increasing $\delta$, 
faster than $\Omega(\D\neq 0,\mu,\delta)$, as can be seen in 
the second curve in fig. (1).
 At some value of $\delta$ the superconducting and the normal phase are both stable. 
The value of $n_u-n_d$ is much higher on the normal phase than in the 
superconducting phase, by the reasons explained above.
One could imagine that, as $n_u-n_d$ increases,  the system proceeds 
towards a mixed phase of $\D=0$ and $\D\neq 0$ phases, with the up-rich $\D=0$ phase
occupying more and more of the space until the superconducting phase completely 
disappears. The presence of two conserved charges  makes things a little 
more involved. The reason is that the total density $n_u+n_d$ 
in the superconducting phase
is much higher than the total density in the normal phase. Thus, even hough
by choosing the right
amounts of the two phases one can arrange so that  the overall asymmetry 
$n_u-n_d$ be any number
between the asymmetries in the two phases, the overall total density is always smaller
than the initial total density. In order to achieve higher values of $n_u-n_d$
and keep $n_u+n_d$ constant we need to increase also the value of the average 
chemical potential $\mu$  (the result in shown on fig. (1), fourth curve to meet the 
 vertical axis, from top to bottom). The value of $\mu$ is chosen in such a way
as to satisfy

\begin{eqnarray}
n_u+n_d &= &x(n_u+n_d)|_{super} + (1-x)(n_u+n_d)|_{normal}\nonumber\\
n_u-n_d &= &x(n_u-n_d)|_{super} + (1-x)(n_u-n_d)|_{normal},
\label{eq:mix2}
\end{eqnarray}
\noindent
for some $x$, $0\le x\le 1$. $\delta$ is determined by the condition that 
$\Omega(\D,\mu,\delta)$
should have two degenerate minima as a function of $\D$ (so phases can coexist), 
so we are left with two equations eq.~(\ref{eq:mix2}) 
and two independent parameters
 ($\mu$ and $x$) so equations eq.~(\ref{eq:mix2}) can, generically,  
be (uniquely) satisfied.
The transition, then, proceeds through a series of mixed states, each 
 one with different values of $\mu$ and $\delta$ 
(arranged in such a way as  to keep $n_u+n_d$ constant while increasing $n_u-n_d$). 
Each one of these mixed states is formed by a slightly asymmetric, high density 
superconducting with a {\it larger} gap than the one before the transition started and
a highly asymmetric, low density normal phase. The mixed phase exists until
its normal component occupies the whole space and the total density of the normal phase
equals the initial value of $n_u+n_d$ (fifth curve to meet the vertical axis in
fig. (1), top to bottom). 
 Increasing $n_u-n_d$ even further is accomplished by increasing $\delta$ again, 
 at roughly fixed $\mu$. At this point the superconducting state is not even 
metastable anymore.

\begin{figure}[t]
\centerline{\epsfxsize=4.0in \epsfbox{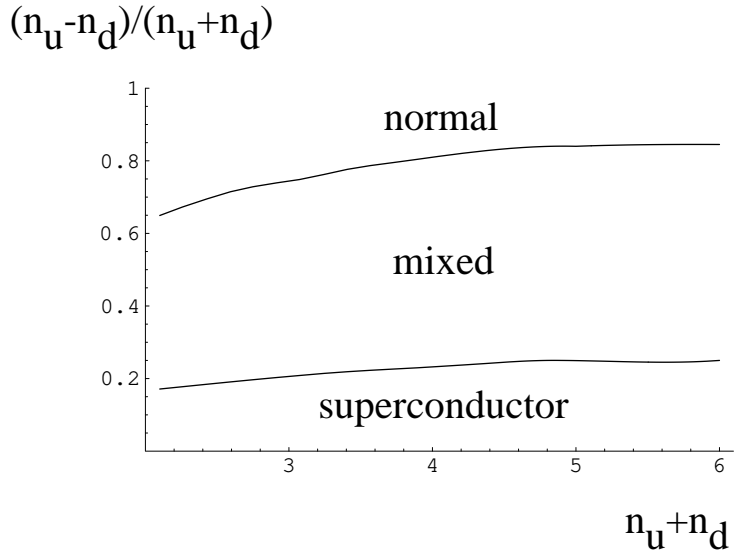}}
\noindent
\caption{\it 
 Phase boundaries as a function of the total density (in units of $10^6$ MeV$^3$. }  
\label{fig:fig2}
\vskip .2in
\end{figure}

The different phases as a function of $n_u+n_d$ and $(n_u-n_d)/(n_u+n_d)$ are sketched
on fig. (2). The horizontal axis, corresponding to the symmetric case, 
is in the superconducting phase for any density, since the BCS instability 
exists for any density. The vertical axis correspond to the completely asymmetric case,
with $\mu_u=-\mu_d$.  There are only up-quarks and down-anti-quarks present, 
no pairing is possible so those states are in the normal phase. 
In between these two phases there  is a large mixed phase region that
 occupy most of the $n_u,n_d\ge 0$ region. The precise location of 
the phase boundaries are somewhat sensitive to the parameters of the model. 
The example shown in fig. (2) was 
obtained with the parameters discussed above. The distance between the 
superconductor-mixed phase is on the large side among those obtained with
other reasonable parameters. As fig. (2) shows, for asymmetries larger than
about $20\%$ the system remains in the mixed phase for arbitrarily high densities.
The phase boundaries in the  low density region, where the model considered here
has no resemblance to real QCD, continue all the way
 to the origin. Unfortunately it is very hard to determine the properties and 
precise location of the mixed phase with a reasonable degree of certainty. 
The bubbles of the normal phase are charged and,
if formed in heavy ion collisions, may lead to charge separation that could be 
observable.


The color-flavor locked state, expected in the 3 massless flavor case, is expected
to behave under asymmetries in a similar manner. Actually, since color is 
completely broken and all quark colors
have a gap, one could expect the mechanisms discussed here to be even more effective
than in the 2SC state. In the case of massive strange quarks though, it remains
to be seen whether existence of ``gapless superconductors'' pointed out in 
\cite{gapless} may hinder this conclusion. A discussion of this
 question, as well as a more complete 
treatment of the two and three flavors asymmetric matter will be presented elsewhere
\cite{next}.

I would like to thank D. Kaplan for discussions and for 
stimulating my interest in the subject. Conversations with S. Reddy and A. Schnell
on related topics are greatly appreciated.

This work is supported in part by the U.S. Dept. of Energy under
Grant No. DOE-ER-40561.


\begin{references}

\bibitem{Barrois}
B. Barrois, Nucl. Phys. {\bf B129} (1977) 390.
S. Frautschi, Proceedings of workshop on hadronic matter at extreme density,
Erice 1978, CALT-68-701.

\bibitem{BailinLove}
D. Bailin and A. Love, Phys. Rept. {\bf 107} (1984) 325.

\bibitem{PisarskiRischke}
R. Pisarski and D. Rischke, {\tt nucl-th/9903023}.

\bibitem{Son}
D. T. Son, Phys. Rev. {\bf D59} (1999) 094019.

\bibitem{Hong}
D. K. Hong, {\tt hep-ph/9812510}; {\tt hep-ph/9905523}.



\bibitem{SW3}
T. Sch\"afer and F. Wilczek, {\tt hep-ph/9906512}. 

\bibitem{PR}
R. Pisarski and D. Rischke, {\tt nucl-th/9907041}.

\bibitem{rockefeller}
W. Brown, J. Liu and H. Ren, {\tt hep-ph/9908248}.

\bibitem{Hsu2}
S. D. H. Hsu and M. Schwetz, {\tt hep-ph/9908310}.
\bibitem{ARW1}
M. Alford, K. Rajagopal and F. Wilczek, Phys. Lett. {\bf B422} (1998) 247.

\bibitem{NJL}  
S.P. Klevansky, Rev. Mod. Phys. {\bf 64} (1992) 649; 
A. Barducci, R. Casalbuoni, G. Pettini and R. Gatto, Phys. Rev. {\bf D49}
(1994) 426.

\bibitem{steph} 
        M. Stephanov, Phys. Rev. Lett. {\bf 76} (1996) 4472;
        Nucl. Phys. Proc. Suppl. {\bf 53} (1997) 469.

\bibitem{RappETC}
R. Rapp, T. Sch\"afer,
E. V. Shuryak and M. Velkovsky, Phys. Rev. Lett. {\bf 81} (1998) 53.

\bibitem{RappETC2}
R. Rapp, T. Sch\"afer,
E. V. Shuryak and M. Velkovsky, {\tt hep-ph/9904353}.

\bibitem{bergesraj} 
J. Berges and K. Rajagopal, Nucl. Phys. {\bf B538} (1999)
215.

\bibitem{stephetal}
M. A. Halasz, A. D. Jackson, R. E. Shrock, M. A. Stephanov
and J. J. M. Verbaarschot, Phys. Rev. {\bf D58} (1998) 096007.

\bibitem{CarterDiakonov}
G. Carter and D. Diakonov, Phys. Rev. {\bf D60} (1999) 016004.

\bibitem{Hsu1}
N. Evans, S. D. H. Hsu and M. Schwetz, Nucl. Phys. {\bf B551} (1999) 275;
Phys. Lett. {\bf B449} (1999) 281.

\bibitem{SW0}
T. Sch\"afer and F. Wilczek, Phys. Lett. {\bf B450} (1999) 325.


\bibitem{CFL}
 M. Alford, K. Rajagopal and F. Wilczek, Nucl. Phys. {\bf B537} (1999) 443.

\bibitem{SW1}
T. Sch\"afer and F. Wilczek,  Phys. Rev. Lett. {\bf 82} (1999) 3956.


\bibitem{gapless}
M. Alford, J. Berges and K. Rajagopal, {\tt hep-ph/9908235}.


\bibitem{next}
P. Bedaque,  in preparation.

\end{references}
\end{document}